\documentclass[aps,pra,showpacs]{revtex4}

\usepackage{graphicx}
\begin{document}

\title{Complex joint probabilities as expressions of determinism in quantum mechanics}

\author{Holger F. Hofmann}
\email{hofmann@hiroshima-u.ac.jp}
\affiliation{
Graduate School of Advanced Sciences of Matter, Hiroshima University,
Kagamiyama 1-3-1, Higashi Hiroshima 739-8530, Japan}
\affiliation{JST, CREST, Sanbancho 5, Chiyoda-ku, Tokyo 102-0075, Japan
}

\begin{abstract}
The density operator of a quantum state can be represented as a complex joint probability of any two observables whose eigenstates have non-zero mutual overlap. Transformations to a new basis set are then expressed in terms of complex conditional probabilities that describe the fundamental relation between precise statements about the three different observables. Since such transformations merely change the representation of the quantum state, these conditional probabilities provide a state-independent definition of the deterministic relation between the outcomes of different quantum measurements. In this paper, it is shown how classical reality emerges as an approximation to the fundamental laws of quantum determinism expressed by complex conditional probabilities. The quantum mechanical origin of phase spaces and trajectories is identified and implications for the interpretation of quantum measurements are considered. It is argued that the transformation laws of quantum determinism provide a fundamental description of the measurement dependence of empirical reality.
\end{abstract}

\pacs{
03.65.Ta, %--Foundations of quantum mechanics; measurement theory; 
03.65.Wj, %--State reconstruction, quantum tomography 
03.67.-a, %--Quantum Information
03.65.Vf %--Phases: geometric; dynamic or topological 
%%%--phase space methods???
}

\maketitle

%%--Introduction
\section{Introduction}
Advances in quantum information technology have established the complete reconstruction of quantum states from experimental data as a standard procedure for the characterization of quantum devices \cite{Leo95,Whi99,Res05,Rie06}. At the heart of these procedures lies the insight that the quantum state of a system with a $d$-dimensional Hilbert space provides a summary of all possible measurement statistics in terms of $d^2-1$ linearly independent elements \cite{Mahler}. It is therefore possible to describe the quantum state in terms of a probability distribution with $d^2$ possible outcomes - the equivalent of a joint probability for the eigenstates of two observables of the system. 

In the light of such complete measurement-based descriptions of quantum states, there has been a growing interest in the identification of fundamental measurement strategies that could serve as a new standard for the evaluation of quantum information encoded in arbitrary states \cite{Bru99,Law02,Pat09,Dia05,Ren04,Fuc09,Med11,Rau09}. Although this research has succeeded in revealing more of the richness of Hilbert space topologies, the formulation of a single standard representation is difficult, since no efficient characterization of quantum states by a discrete set of measurement operators can reflect the continuous symmetry of Hilbert space with regard to unitary transformations between different measurements. It seems that all attempts to identify fundamental measurements must necessarily introduce a bias that is not found in the original Hilbert space formalism with its equivalent representation of all projective measurements as orthogonal basis systems of an isotropic vector space. However, the conventional Hilbert space representation of quantum statistics in terms of a single orthogonal basis set is even more biased, as it represents only half of the physics in terms of measurement results, while the other half is encoded in terms of abstract quantum coherences. Since quantum coherences of one basis show up as probabilities in another basis, it may be desirable to find a more symmetric description of the density operator that expresses the coherences of a quantum state in terms of joint probabilities for non-orthogonal measurement outcomes.

Since the number of elements needed to describe the complete density operator is exactly equal to the number of eigenstate combinations of two observables, it would seem natural to represent the quantum state as a joint probability of only two measurements. Intuitively, any pair of basis sets with non-zero mutual overlap should represent two independent pieces of information distinguishing $ d\times d = d^2$ elements of the statistics. Different choices of observables would then correspond to different parameterizations of the two-dimensional phase space topology defined by any pair of observables with mutually overlapping eigenstates. Interestingly, such a joint probability was already proposed very early in the history of quantum mechanics, as an alternative to the Wigner function in phase space \cite{Kir33,Joh07}. Essentially, this joint probability is obtained by multiplying the projection operators for the two measurements and taking the expectation value of the resulting complex-valued operator. It is therefore the most natural definition of joint probability that the quantum formalism provides for measurements that cannot actually be performed jointly. 

One of the reason that Kirkwood's approach to joint probabilities in quantum mechanics received disappointingly little attention may be the lack of practical applications of the theory. After all, the Copenhagen interpretation of quantum mechanics implies that questions about events that do not happen are inherently meaningless. However, quantum paradoxes clearly show that quantum mechanics makes non-trivial statements about the statistical relations between measurements that cannot be performed at the same time. In fact, the paradoxes show that quantum mechanics cannot be understood in terms of positive joint probabilities for measurement outcomes represented by non-commuting measurement operators. Recently, it has been shown that the paradoxical aspects of quantum statistics are consistent with negative conditional probabilities determined in weak measurements
\cite{Aha88,Mir07,Lun09,Yok09,Gog11,Lun11}. The consistency of these results strongly suggests that the statistics of weak measurements is a fundamental element of the Hilbert space formalism. In particular, it is possible to develop a consistent explanation of weak measurement statistics in terms of complex conditional and joint probabilities \cite{Ste95,Hof10,Hos10,Hof11a}. 

Interestingly, the joint probabilities derived from weak measurement are identical with the joint probability originally introduced by Kirkwood on purely mathematical grounds \cite{Joh07}. It is therefore possible to express any quantum state in terms the joint probability distribution obtained from the weak measurement of $\mid a \rangle \langle a \mid$, followed by a final measurement of $\mid b \rangle \langle b \mid$, where $\langle a \mid b \rangle \neq 0$. Alternatively, weak measurement statistics can also be obtained from the measurement back-action of projective measurements \cite{Johx}, or from the correlations between optimal quantum clones of the input state \cite{Hofx}. Complex joint probabilities thus provide a surprisingly consistent description of the correlations between pairs of measurements that cannot be performed jointly. 

However, there remains an important question that needs to be addressed: complex probabilities cannot be interpreted as relative frequencies of microscopic realities. Therefore, they cannot be identified with classical phase space points. In particular, the measurement outcomes for a third measurement $c$ cannot be related to well-defined pairs of measurement outcomes $(a,b)$, as the classical phase space analogy would suggest. Nevertheless, a description of the quantum state in terms of joint probabilities for $b$ and $c$ is just as complete as a description based on $a$ and $b$. Therefore, the transformation between the two representations is reversible and deterministic. In the following, I will take a closer look at this relation between different joint probabilities. It is shown that the deterministic transformation is given by the complex conditional probabilities $p(c|a,b)$ that characterize weak measurement statsitics \cite{Hof11a}. Reversibility of the transformation requires that the information about $a$ can be recovered completely from the information about $c$ after the transformation. For positive probabilities, this condition requires that $c$ is a well-defined function of $a$ and $b$. In quantum mechanics, the same mathematical relation is fulfilled as a result of the orthogonality of the Hilbert space vectors $\{\mid a \rangle\}$. The structure of Hilbert space can then be understood as a modification of determinism that reconciles continuous transformations with discrete measurement results at the expense of microscopic realism. 

The fact that the fundamental expression of determinism in quantum mechanics can be represented by complex conditional probabilities has significant implications for the formulation of the classical limit that represents the conventional notion of determinism in physics. Specifically, this classical determinism only emerges as a macroscopic approximation to the microscopic quantum description. To illustrate this emergence of classical realism, it is necessary to introduce the concept of measurement resolution, based on a sequence of quantum states that defines the distance between two orthogonal states. With this metric, the complex phases of the conditional probabilities $p(c|a,b)$ can be identified with phase space distances \cite{Hof11a}. Since large phase space distances correspond to rapid phase oscillations in $c$, coarse graining rapidly reduces the precise expression of quantum determinism in terms of complex probablities to a single-peaked function centered around a single value of $c$, as expected from classical determinism. Quantum determinism can thus explain how the classical notion of a measurement independent reality emerges as an approximation to the more accurate description of context dependent realities in quantum mechanics. 

The rest of the paper is organized as follows. In section \ref{sec:jointprob}, the representation of quantum states as complex joint probabilities of observables with mutually overlapping eigenstates is introduced and the operator algebra is defined. In section \ref{sec:transform}, it is shown that transformations between different measurements are expressed by the complex conditional probabilities corresponding to the weak values of the projection operators for the new basis. The general criterion for quantum determinism is derived and the differences between classical determinism and quantum determinism are discussed. In section \ref{sec:topology}, it is shown that a classical phase space topology emerges in higher dimensional Hilbert spaces. Quantum determinism is still fundamentally different from classical determinism, but they become indistinguishable when the resolution of a measurement result is limited by Gaussian noise. Measurement independent phase space points therefore emerge as approximate realities in the limit of low measurement resolution. In section \ref{sec:causality}, quantum determinism is applied to unitary dynamics and different representations of causality are considered. It is pointed out that the identification of quantum dynamics with paths or histories described by a sequence of measurement results may be a misinterpretation of quantum determinism based on the extrapolation of realist notions beyond their natural limit of validity. In section \ref{sec:complex}, it is pointed out that the imaginary part of complex joint probability does have a classical limit, represented by the gradients of the classical phase space distribution. Quantum corrections to classical determinism become relevant when the imaginary part of the joint probability becomes comparable to the real part. In section \ref{sec:discuss}, the empirical foundations of quantum determinism are reviewed and consequences for the interpretation of quantum mechanics are considered. It is emphasized that complex joint probabilities do not represent relative frequencies of quasi-realities, but should be understood as the fundamental deterministic relations between measurements that can never be performed jointly. Quantum determinism therefore highlights the dependence of empirical reality on the measurement context. 

%%-------------------

\section{Joint probability representation of quantum states}
\label{sec:jointprob}

A complete description of quantum statistics in terms of measurement probabilities is not a straightforward matter, because the uncertainty principle generally prevents the joint performance of separate quantum measurements. It is therefore impossible to simultaneously measure two observables with different eigenstates. In principle, it is possible to perform a sequence of measurements on the same system, but then the measurement interaction of the first measurement will change the result of the second measurement, so that the outcome of the second measurement cannot be identified with the value of the observable before the first measurement. 

Interestingly, there exist situations where sequential measurements can be interpreted as joint measurements. This is the limit of weak measurements \cite{Aha88}, where the measurement interaction of the first measurement is so low that its effect on the second measurement is negligible. Although the signal-to-noise ratio of weak measurements is much smaller than one, the average measurement results are consistent with the expectation values of the measured observables. A final measurement can then identify the conditional expectation values, also known as weak values. 

The complex joint probabilities obtained from weak measurements have a particularly simple mathematical form. In general, they correspond to the expectation value obtained for the operator product of the two measurement operators \cite{Hof10}. For two observables with mutually overlapping sets of eigenstates $\{\mid a \rangle\}$ and $\{\mid b \rangle\}$, the complex joint probabilities representing the density operator $\rho$ of an arbitrary quantum state are therefore given by the expectation value of the ordered product of the projection operators $\mid b \rangle \langle b \mid$ and $\mid a \rangle \langle a \mid$,
\begin{equation}
\label{eq:jointprob}
\rho(a,b)=\langle b \mid a \rangle \langle a \mid \hat{\rho} \mid b \rangle.
\end{equation}
As pointed out by Johansen \cite{Joh07}, this is identical to the joint probability introduced by Kirkwood in 1933 \cite{Kir33}. Johansen also showed that the complex probabilities provide a complete expansion of the density operator, with very convenient mathematical properites \cite{Joh07}. 

An essential advantage of the joint probability representation given in Eq.(\ref{eq:jointprob}) is that it stays very close to the original Hilbert space formalism, where the density matrix is defined in terms of a single measurement basis. Effectively, the joint probability can be understood as a partial transformation of the right side of the $a$-basis density matrix to the $b$-basis, followed by an adjustment with a complex overlap factor of $\langle b \mid a \rangle$. This transformation is obviously reversible for all transformations with non-zero overlap $\langle b \mid a \rangle$. The complex joint probability of $a$ and $b$ thus provides a complete expression of quantum coherences without the need for interferences between mutually exclusive alternatives.  

Classical joint probabilities refer to joint realities of $a$ and $b$. In the quantum formalism, this corresponds to a normalized contribution to the density operator with simultaneous probabilities of one for both $\mid a \rangle$ and $\mid b \rangle$. For the complex joint probabilities of Eq.(\ref{eq:jointprob}), this set of basis operators is given by
\begin{equation}
\hat{\Lambda}(a,b) = \frac{\mid a \rangle\langle b \mid}{\langle b \mid a \rangle}. 
\end{equation} 
The operators $\Lambda(a,b)$ are orthogonal with regard to the adjoint product trace, where the norm of the operators is given by the inverse overlap of $\mid a \rangle$ and $\mid b \rangle$,
\begin{equation}
\mbox{Tr}\left(\hat{\Lambda}(a,b)\hat{\Lambda}^\dagger(a^\prime, b^\prime)\right)=\frac{1}{|\langle b \mid a \rangle|^2}
\delta_{a,a^\prime}\delta_{b,b^\prime}.
\end{equation} 
Using this $d^2$-dimensional orthogonal operator basis, any density operator can be expressed in terms of the complex joint probabilities of $a$ and $b$,
\begin{eqnarray}
\label{eq:expand}
\hat{\rho} &=& \sum_{a,b} |\langle b \mid a \rangle|^2  \mbox{Tr}\left(\hat{\rho} \hat{\Lambda}^\dagger(a, b)\right)
\hat{\Lambda}(a,b) 
\nonumber \\
&=& \sum_{a,b} \rho(a,b) \; \hat{\Lambda}(a,b).
\end{eqnarray}

Eq.(\ref{eq:expand}) shows that complex joint probabilities are a complete representation of quantum statistics, regardless of measurement context. It is therefore possible to represent all measurement statistics in terms of the statistics relating to the measurements of $a$ and $b$. In particular, the expectation values of all self-adjoint operators $\hat{M}$ can be defined in terms of $a$ and $b$ by simply expanding the operators in the adjoint operator basis $\{\Lambda^\dagger(a,b)\}$. The coefficients of this expansion are given by 
\begin{equation}
\mbox{Tr}\left(\hat{\Lambda}(a,b)\hat{M}\right) = \frac{\langle b \mid \hat{M} \mid a \rangle}{\langle b \mid a \rangle}.
\end{equation}
Since these are the weak values of the operator $\hat{M}$ observed for an initial state of $\mid a \rangle$ and a final state of $\mid b \rangle$, the complex joint probabilities $\rho(a,b)$ appear to describe the density matrix as a mixture of transient quantum states $\{\hat{\Lambda}(a,b)\}$ defined by the respective combinations of initial and final states \cite{Hof10,Shi10}. Consequently, the expectation value of $\hat{M}$ corresponds to the average weak value given by 
\begin{equation}
\label{eq:expect}
\langle \hat{M} \rangle = \sum_{a,b} \rho(a,b) \; \frac{\langle b \mid \hat{M} \mid a \rangle}{\langle b \mid a \rangle}.
\end{equation}
In the light of the formal similarity to classical statistics, it may be important to remember that this expectation value describes the results of a direct measurement of $\hat{M}$, and not the results of weak measurements. The weak value of $\hat{M}$ is therefore not just an experimental result, but also a fundamental element of the operator algebra, similar to the values of operators obtained for a point in phase space in the Wigner transformation of an operator. The complex weak values of $\hat{M}$ conditioned by $a$ and $b$ thus provide a complete mathematical expression of the operator $\hat{M}$.

The possibility of constructing joint probability representations for nearly arbitrary pairs of observables raises a few interesting questions about the relation between observables and the structure of Hilbert space. Specifically, any pair of observables can now serve as a ``parameterization'' of quantum states. If the results could be interpreted in terms of classical joint probabilities, each pair of values $(a,b)$ would designate a microstate defining a phase space point.
Keeping this analogy in mind, the transformation between different measurement bases corresponds to a change of coordinates in the effective phase space. In the following, I will examine the quantum mechanical expressions that describe such transformations in the extreme quantm limit.

%---------

\section{Transformations of complex joint probabilities}
\label{sec:transform}

Complex joint probabilities can be formulated for any pair of observables whose eigenstates have non-zero mutual overlap. It is therefore possible to transform complex joint probabilities between different basis sets representing different measurements. If we consider the transformation from a basis set $(\{\mid a \rangle\},\{\mid b \rangle\})$ to a basis set  $(\{\mid c \rangle\},\{\mid b \rangle\})$, the transformation is given by 
\begin{equation}
\label{eq:transform}
\rho(c,b) = \sum_a p(c|a,b) \rho(a,b),
\end{equation}
where the coefficients of the transformation are given by the weak conditional probabilities $p(c|a,b)$ with
\begin{equation}
\label{eq:condprob}
p(c|a,b) = \frac{\langle b \mid c \rangle \langle c \mid a \rangle}{\langle b \mid a \rangle}.
\end{equation}
These conditional probabilities are equal to the weak values of the projection operators of $\mid c \rangle$ and thus correspond to the probability of finding $c$ conditioned by an initial value of $a$ and a final value of $b$. A statistical interpretation of this transformation would suggest that the relation between $a$ and $c$ is random, corresponding to an irreversible scattering of inputs $a$ into different outputs $c$. However, the transformation is merely a change of representation and does not change the physical properties of the state. It should therefore be fully deterministic. 

A formal definition of determinism can be obtained from the reversibility of the transformation. If the transformation from $a$ to $c$ is deterministic, the original joint probability can be recovered by the inverse transformation represented by the conditional probabilities $p(a^\prime|c,b)$. Therefore, conditional probabilities can only describe a deterministic transformation between different representations of the same probability distribution if they satisfy the relation
\begin{equation}
\label{eq:determinism}
\sum_c p(a^\prime|c,b) p(c|a,b) = \delta_{a,a^\prime}.
\end{equation} 
For classical statistics, where probabilities are real and positive, the above relation can only be satisfied if the conditional probabilities assign a specific value of $c$ to each value of $a$, so that the conditional probabilities are one for the correct assignment and zero for all other assignments. In the quantum limit, the relation is still valid, but instead of taking only values of zero or one, the complex conditional probabilities reflect the structure of Hilbert space, as shown by the contributions from each value of $c$,
\begin{equation}
\label{eq:Hilbert}
p(a^\prime|c,b) p(c|a,b) = \frac{\langle b \mid a^\prime \rangle}{\langle b \mid a \rangle}
\langle a^\prime \mid c \rangle \langle c \mid a \rangle.
\end{equation}
Thus, the quantum limit of determinism is obtained from the orthogonality of $\mid a \rangle$ and $\mid a^\prime \rangle$, even though there is no conditional assignment of a fixed value of $c$ to each pair of values $(a,b)$. In fact, quantum determinism as defined by the conditional probabilities in Eq. (\ref{eq:condprob}) not only fails to assign a specific value of $c$ to each pair $(a,b)$, but actually assigns a non-zero value to the complex probability of each state $\mid c \rangle$ that is not orthogonal to either $\mid a \rangle$ or $\mid b \rangle$. For basis sets with non-zero mutual overlap, the relation between $c$ and $(a,b)$ is therefore spread out over all possible combinations of $a$, $b$, and $c$. Determinism only emerges because of the complex phases of the conditional probabilities. 

To recognize the significance of the difference between classical determinism and quantum determinism, it is useful to consider the classical interpretation of joint probabilities as relative frequencies of microstates defined by the phase space point $(a,b)$. In this case, deterministic transformations can only correspond to an exchange of labels denoting the fundamental representation independent reality of the phase space point $(a,b)=(c,b)$. On the other hand, quantum determinism prevents the identification of such representation independent realities. The statistical relations defined by the Hilbert space structure of quantum mechanics imply that the mathematical points $(a,b)$ are fundamentally different from the mathematical points $(c,b)$. Quantum determinism is therefore completely detached from classical realism. In the next section, I will illustrate the transition between quantum determinism and classical determinism by constructing a phase space over a sufficiently large Hilbert space. It is then possible to see how the classical notion of a measurement independent reality can emerge as an approximation of the more accurate relations of contextual quantum determinism. 

%-----------

\section{Emergence of phase space topologies from quantum determinism}
\label{sec:topology}

A fundamental contradiction between classical determinism and quantum determinism arises in discrete systems, where quantum determinism allows continuous transformations, whereas classical determinism only allows discrete exchanges of points. As a result, it is difficult to construct a phase space topology for few level systems. Even in the limit of high dimensional Hilbert spaces, it is not immediately clear how to identify quantum states with parameters. In practical systems, this parameterization usually emerges from the interactions with the environment, which introduces a sequence of states, so that the distance between two orthogonal states $\mid a \rangle$ and $\mid a^\prime\rangle$ can be expressed as a numerical difference of $a-a^\prime$. Continuous phase space topologies then emerge when the discrete steps of $\pm 1$ in $a$ or $b$ can be considered microscopically small. For basis sets with non-zero overlap, the conditional probabilities $p(c|a,b)$ can then be given by continuously varying functions of $a$, $b$ and $c$. If the absolute values of the overlaps between the states vary only slowly, the phase of the complex conditional probability can be expanded in a Taylor series up to second order around an extremum, resulting in a complex Gaussian with an imaginary variance of $i V_q$,
\begin{equation}
\label{eq:phase}
p(c|a,b)= \frac{1}{\sqrt{2 \pi V_q}} \exp(i \frac{(c-f_c(a,b))^2}{2 V_q} - i \frac{\pi}{4}).
\end{equation}
Since the phase also varies slowly in $a$ and $b$, $f_c(a,b)$ can be approximated by a linear function of $a$ and $b$. Comparison with Eq.(\ref{eq:condprob}) shows that the imaginary variance is given by
\begin{equation}
V_q = \frac{|\langle b \mid a \rangle|^2}{2 \pi |\langle b \mid c \rangle|^2 |\langle c \mid a \rangle|^2}.
\end{equation}
The gradients of $f_c(a,b)$ can be determined by considering the normalizations of $p(a|c,b)$ and $p(b|a,c)$. The results read
\begin{eqnarray}
\frac{\partial}{\partial a} f_c(a,b) &=& \frac{|\langle b \mid a \rangle|^2}{|\langle b \mid c \rangle|^2}
\nonumber \\
\frac{\partial}{\partial b} f_c(a,b) &=& \frac{|\langle b \mid a \rangle|^2}{|\langle c \mid a \rangle|^2}.
\end{eqnarray}
The conditional probability $p(c|a,b)$ is therefore completely determined by the Hilbert space overlaps of the basis states. At the same time, $a$, $b$ and $c$ correspond to phase space coordinates, where $f_c(a,b)$ defines the corresponding classical coordinate transformation. 

\begin{figure}[th]
\begin{picture}(500,320)
\put(0,0){\makebox(480,320){\vspace*{-4cm}
\scalebox{0.9}[0.9]{
\includegraphics{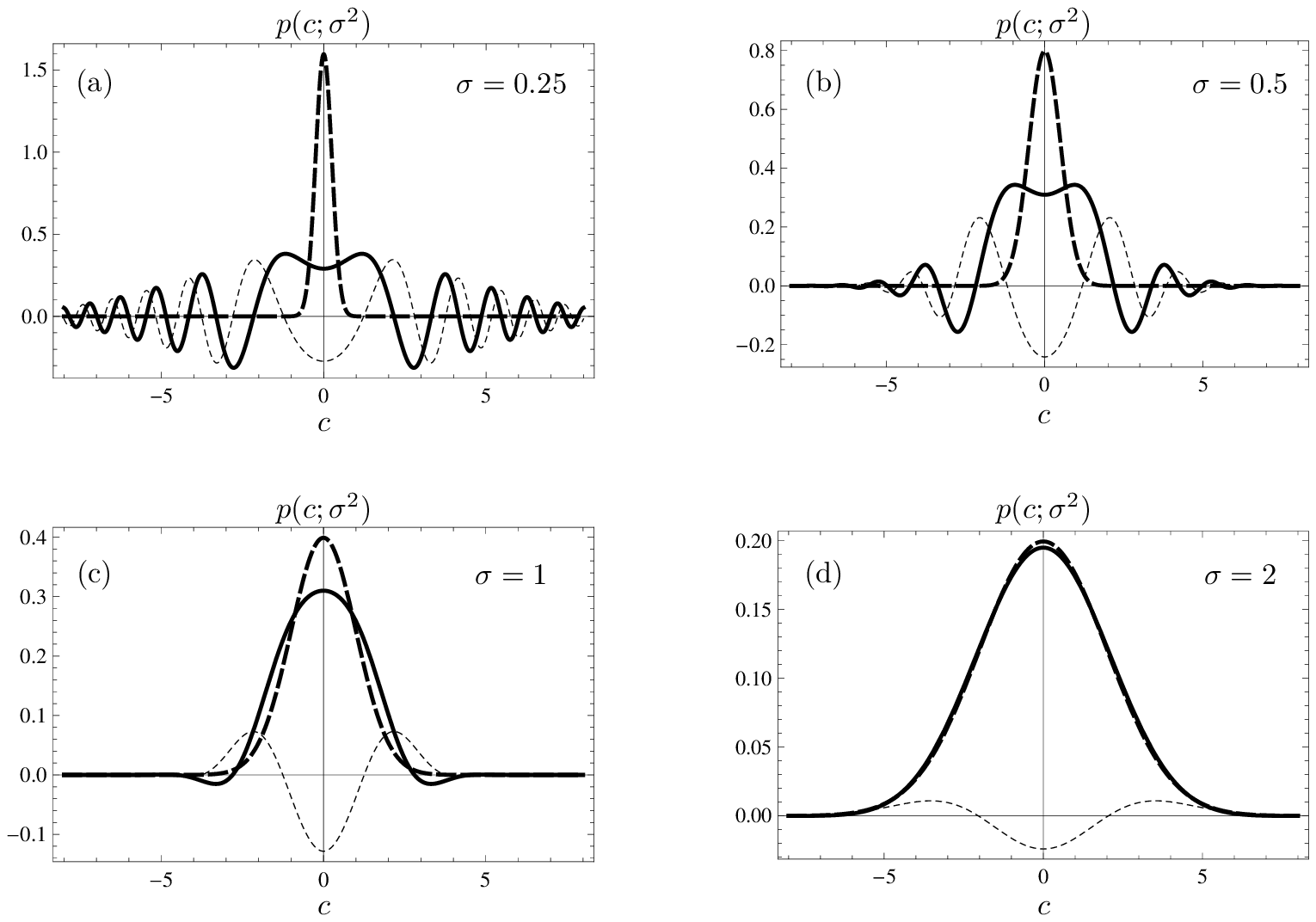}}}}
\end{picture}
\caption{\label{fig1} Comparison between the complex conditional probabilities of quantum determinism and the corresponding classical predictions for an imaginary variance of $V_q=1$ in $c$ for different Gaussian resolutions $\sigma$. Thick lines show the real part of the complex probability $p(c;\sigma^2)$, thick dashed lines show the corresponding classical probability distribution, and thin dashed lines show the imaginary part of $p(c;\sigma^2)$. (a) illustrates the difference between quantum determinism and classical predictions at a high resolution of $\sigma=0.25$, (b) shows the transition to low resolution at $\sigma=0.5$, (c) shows the similarity of quantum statistics and classical statistics at $\sigma=1$, and (d) shows the small deviations from the classical limit that remain at $\sigma=2$.}
\end{figure}

Since quantum determinism requires that the absolute values of all conditional probabilities are non-zero, it is fundamentally different from classical determinism, where conditional probabilities of zero are assigned to all combinations of $a$, $b$ and $c$ that do not fulfill the functional dependence given by $c=f_c(a,b)$. Instead, quantum determinism represents the relation between $a$, $b$ and $c$ in terms of complex phase oscillations. Specifically, the functional dependence given by $f_c(a,b)$ defines the values of $c$ for which the complex phase of $p(c|a,b)$ achieves its minimum. Classical realism emerges if this phase minimum can be identified with the only relevant value of $c$. In this case, $p(c|a,b)$ can be replaced by a delta function, $\delta(c-f_c(a,b))$. To see how well classical realism can approximate the more precise quantum results, it is possible to compare the predictions of quantum determinism and classical realism for coarse grained probabilities, e.g. by folding the conditional probabilities $p(c|a,b)$ with a Gaussian of variance $\sigma^2$. For the classical probability $\delta(c-f_c(a,b))$, the result is a Gaussian with variance $\sigma^2$ around $(c-f_c(a,b))$. The precise result obtained from the complex conditional probability in Eq.(\ref{eq:phase}) can be written as
\begin{equation}
\label{eq:climit}
p(c;\sigma^2) = \frac{1}{\sqrt{2 \pi \sigma^2(1+i\epsilon)}} \exp\left(\frac{(c-f_c(a,b))^2}{2 \sigma^2 (1+\epsilon^2)}(1-i \epsilon)\right),
\end{equation}
where $\epsilon=V_q/\sigma^2$ describes the relative deviation from the classical probability distribution. Clearly, quantum determinism converges on classical determinism for small values of $\epsilon$. This means that quantum determinism is indistinguishable from classical determinism at resolutions of $c$ much lower than $\sqrt{V_q}$. Fig. \ref{fig1} illustrates this rapid disappearance of experimentally observable contradictions between the predictions of classical realism and quantum determinism. Since the low resolution limit characterizes almost all of our actual experience, our intuitive notion of realism may well be explained as a product of this classical approximation to quantum determinism.

%%--figure 

In addition to the classical functional relations $c=f_c(a,b)$ that relate different parameterizations of phase space to each other, classical phase space also has a well-defined metric that ensures the conservation of phase space volume under all canonical transformations. In quantum mechanics, this metric corresponds to the density of states in the phase space volume defined by changes of $a$ and $b$. In the discussion above, $a$ and $b$ are integers that number the discrete basis states of a $d$-dimensional Hilbert space. For this quantum mechanical parameterization, the metric of phase space emerging in the classical limit is found by replacing the sum over all values of $a$ and $b$ with approximate integrals, so that the total number of states is given by
\begin{equation}
\int_1^d \int_1^d |\langle a \mid b \rangle|^2 \,da \; db = \int_1^d db \approx d.
\end{equation}
The metric emerging from a derivation of phase space from a Hilbert space parametrized by numbering the states therefore has a metric that is given in terms of the density of quantum states, which is equal to $|\langle a \mid b \rangle|^2$ near the phase space point $(a,b)$. A canonical parameterization of phase space is obtained for 
\begin{equation}
\langle a \mid b \rangle = \frac{1}{\sqrt{d}} \exp\left(i \frac{2 \pi}{d} \, a b \right).
\end{equation}
In this case, unitary phase shifts in $a$ generates shifts in $b$, and vice versa. The parameters $a$ and $b$ can be re-scaled in units of position $x$ and momentum $p$, so that the phase of $\langle x \mid p \rangle$ is given by $x p/\hbar$. This re-scaling shows how the classical action emerges from the quasi-continuous limit of joint probability representations in sufficiently large Hilbert spaces. 

%----

\section{Causality as quantum determinism}
\label{sec:causality}

According to classical causality, a single point in phase space defines the properties of a closed system at all times. In this sense, the canonical phase space coordinates of position and momentum can be interpreted as parametrization based on a specific reference time, and the time evolution of the coordinates represents transformations to different parameterizations of the same phase space. In general, it is therefore possible to define phase space parameterizations referring to multiple times and even to weighted averages over time.  

In classical determinism, this ambiguity of phase space concepts is not particularly relevant, since it is always possible to identify the continuous time-evolution of observable properties in terms of well-defined time-dependent functions. However, the situation is quite different in the limit of quantum determinism. Here, simultaneous statements about the same property at different times do not usually commute. Therefore, it is not correct to assign reality to a continuous trajectory describing the dynamics of the system. It may well be the case that the focus on dynamics and time evolution in traditional physics has unnecessarily complicated the picture we have of quantum mechanics. Quantum determinism addresses this problem by describing the time evolution of closed systems as a re-parameterization of an unchanged quantum state $\hat{\rho}$. Causality in quantum mechanics is then described by the complex conditional probabilities of quantum determinism for statements associated with different times. 

The conventional representation of deterministic causality in quantum mechanics is given by the unitary transformation $\hat{U}(t_j-t_i)$ that defines the relation between states at time $t_j$ with states at time $t_i$. If a quantum state $\hat{\rho}$ is expressed by the complex joint probability $\rho(a_1,b_1)$ of the properties $a_1$ and $b_1$ at time $t_1$, the transformation to $a_2$ and $b_2$ at time $t_2$ should proceed in two steps, since elementary quantum determinism describes the relations between sets of three observables. For example, determinism defines the value of $a_2$ at time $t_2$ as a function of both $a_1$ and $b_1$. Therefore, either $a_1$ or $b_1$ can be replaced by $a_2$. For reasons of symmetry, the natural choice seems to be a transformation to $(a_1,a_2)$,
\begin{equation}
\rho(a_1,a_2) = \sum_{b_1} p(a_2|a_1,b_1) \rho(a_1,b_1).
\end{equation}  
This two-time representation of the quantum state reflects the fact that trajectories can be defined by the positions at two different times. Since this representation is in principle equivalent to any other, the evolution of $a_i$ up to a third time $t_3$ can be evaluated directly from the complex joint probabilities of $a_1$ and $a_2$,
\begin{equation}
\rho(a_1,a_3) = \sum_{a_2} p(a_3|a_1,a_2) \rho(a_1,a_2).
\end{equation}
Here, the conditional probability $p(a_3|a_1,a_2)$ corresponds to the probability of finding the system in $a_3$ at time $t_3$, when it was initially in $a_1$ and finally arrived in $a_2$. For the positions of a free particle, the complex phase of this conditional probability is given by the action of the trajectory $a_1 \to a_3 \to a_2$, so that the classical result for $a_3$ corresponds to the path of least action \cite{Hof11a}. 

The connections between extended probabilities and path integrals or quantum histories have already been noted in other works \cite{Sok06,Har08}. However, the explanations given there seem to be at odds with determinism, since the representations appear to assign a non-deterministic time evolution to a single quantum object. Nevertheless quantum determinism can reproduce the same results in terms of a gradual transformation from $a_2$ at $t_2$ to $a_n$ at $t_n$ in a number of steps evolving $a_i$ at $t_i$ to $a_{i+1}$ at $t_{i+1}$. The total conditional probability for the transformation is then given by
\begin{eqnarray}
\label{eq:path}
p(a_n|a_1,a_2)&=&\sum_{\{a_i\}} p(a_n|a_1,a_{n-1})\,p(a_{n-1}|a_1,a_{n-2})\;\ldots
\nonumber \\ && \hspace{4cm} \ldots \;
p(a_4|a_1,a_3)\,p(a_3|a_1,a_2),
\end{eqnarray} 
which converges on the path integral for the evolution of $a(t)$ in the limit of continuous times. Specifically, the phase of each contribution to the sum over the paths $\{a_i\}$ is defined by a sum corresponding to the total action of that path. Since sums over rapidly oscillating phases cancel out, the end result can be obtained by summing over only a finite interval around the classical trajectory given by the path of least action. 

Although Eq.(\ref{eq:path}) shows that the conditional probabilities of quantum determinism can be expressed in terms of path integrals, it seems significant that these path integrals do not describe the evolution of a quantum state. Instead, they describe a sequential transformation of state-independent conditional probabilities that describe the fundamental deterministic relations between the non-commuting observables $a_i$.
Quantum determinism thus provides an alternative explanation for the role of path integrals in the description of the dynamics of a system. Specifically, the transformations in Eq.(\ref{eq:path}) are merely a change of representation. It is therefore difficult to justify the interpretation of an individual path as the history of an individual system, even though the formal assignment of a complex probability to each path is indeed possible \cite{Sok06,Har08}. Clearly, each path is merely a sequence of statements, each of which can be translated to equivalent statements at other times. Since a pair of statements is in principle sufficient to define the statistics of all other statements, the paths are merely redundant representations of the fully deterministic evolution of the physical properties that characterize the system. The misleading impression that a quantum system could ``choose'' between alternate paths or histories arises from a misinterpretation of joint probabilities with joint realities. As we saw in the previous section, such an identification represents an approximation valid only in the classical limit of low measurement resolution. 

In Hilbert space, the time evolution of quantum states is represented by unitary transformations $\hat{U}(t_j-t_i)$ generated by the Hamilton operator $\hat{H}$. If only the time evolution of a single measurement outcome $a$ is of interest, it may therefore be convenient to express the quantum state as a complex joint probability of $\mid a(t)\rangle = \hat{U}(t) \mid a \rangle $ and an eigenstate $\mid n \rangle$ of the Hamiltonian $\hat{H}$ with an energy eigenvalue of $E_n$. The time evolution can then be expressed in terms of the complex conditional probability $p(a(t)|n,a^\prime)$. The time dependence of this conditional probability corresponds to the formulation of the time dependent Schroedinger equation in the $\{\mid a \rangle\}$-basis,
\begin{eqnarray}
\label{eq:schro}
\lefteqn{\frac{d}{dt} (p(a(t)|n,a^\prime)\; \langle a^\prime\mid n \rangle)
=}
\nonumber \\ && \hspace{1cm}
 -\frac{i}{\hbar} \sum_{a^{\prime\prime}} \langle a^\prime \mid (\hat{H}-E_n) \mid a^{\prime\prime} \rangle 
\;(p(a(t)|n,a^{\prime\prime}) \;\langle a^{\prime\prime}\mid n \rangle).
\end{eqnarray}
Essentially, the re-scaled conditional probabilities $p(a(t)|n,a^\prime) \langle a^\prime\mid n \rangle$ evolve just like the $a^\prime$-components of a state vector. In the limit of smoothly varying phases, these dynamics therefore correspond to the well known dynamics of dispersion in wave propagation. Quantum determinism thus reproduces the formal aspects of the wave-particle dualism implied by the conventional formulation of the Schroedinger equation. However, the re-formulation in terms of conditional probabilities for measurements at different times shows that the object of the dynamical evolution is not a physical wave, but the statistics of statements about a property of the quantum system at different times. The deeper meaning of the formal analogy between the elastic properties of physical waves and the conditional statistics of post-selected measurements is therefore far from obvious, and related measurement results such as \cite{Lun11} should not be misinterpreted in terms of a ``realism'' of the wavefunction.

The analysis of Hamiltonian dynamics also reveals a highly non-classical relation between transformation dynamics and statistics that can be expressed in the form of complex probabilities \cite{Hof11a}. In its most simple form, this relation is expressed by the definition of imaginary weak values as logarithmic derivatives of the post-selected probabilities for a weak unitary transformation generated by the respective observable \cite{Hof11b}. The time evolution of measurement probabilities can therefore be expressed in terms of imaginary weak values of energy, 
\begin{eqnarray}
\label{eq:Imweak}
\frac{d}{dt} \langle a \mid \hat{\rho} \mid a \rangle &=& - \frac{i}{\hbar}\left(\langle a \mid \hat{H} \hat{\rho} \mid a \rangle - \langle a \mid \hat{\rho} \hat{H} \mid a\rangle \right)
\nonumber \\ &=& 
\sum_n \frac{2 E_n}{\hbar} \; \mbox{Im}\left(\rho(E_n,a)\right).
\end{eqnarray} 
This expression provides a direct interpretation of imaginary probabilities that is consistent with classical theories of phase space transformations. In the following, I will use this analogy to provide a classical definition of complex probability that corresponds to the low resolution limit of the quantum mechanical values. 

%-------

\section{Complex probability in the classical limit}
\label{sec:complex}

As shown in section \ref{sec:topology}, classical phase space features emerge as soon as Hilbert space is sufficiently large to allow a representation of quantum phases and amplitudes as smooth continuous functions of the variables $a$ and $b$. for a discussion of classical limits, it is therefore often sufficient to focus on a continuous variable phase space defined in terms of position $\hat{x}$ and momentum $\hat{p}$. The complex joint probability of a quantum state is then given by
\begin{equation}
\rho(x,p)=\langle p \mid x \rangle \langle x \mid \hat{\rho} \mid p \rangle,
\end{equation}
where $\langle p \mid x \rangle = \exp(-i px/\hbar)/\sqrt{2 \pi \hbar}$. Incidentally, this is precisely the form in which Kirkwood originally introduced the complex probability distribution in 1933, as an alternative to the Wigner function \cite{Kir33}. However, it gained much less recognition than the Wigner function, probably mostly because the complex phases appear to complicate the comparison with classical statistics. It is therefore a somewhat ironic twist that the Kirkwood distribution actually describes the measurement statistics observed in weak measurements, to the point where its discrete versions can resolve quantum paradoxes. The Kirkwood distribution thus provides the correct continuous variable limit of the more general discrete quantum statistics that can be observed and verified by weak measurements. 

In general, the imaginary part of complex probabilities can be defined operationally as logarithmic derivatives of measurement probabilities in response to weak transformations generated by the observable in question \cite{Hof11b}. In particular, Eq.(\ref{eq:Imweak}) shows how the time evolution of a measurement distribution depends on the imaginary parts of the joint probability with the eigenstates of the Hamiltonian. This relation can be applied to position and momentum by considering the change in a momentum distribution $\rho(p)$ caused by a potential $V(x)$,
\begin{equation}
\label{eq:Imxp}
\frac{d}{dt} \rho(p) = \frac{2}{\hbar} \int V(x) \mbox{Im}\left(\rho(x,p)\right) dx.
\end{equation}
In the classical limit, the change of momentum is given by $dp/dt=-\partial V/\partial x$, so the relation between the change of $\rho(p)$ and the real-valued joint probability reads
\begin{equation}
\label{eq:Rexp}
\frac{d}{dt} \rho(p) = \int \frac{\partial}{\partial x} V(x) \frac{\partial}{\partial p} \mbox{Re}\left(\rho(x,p)\right) dx.
\end{equation}
Integration in parts can be used to identify the imaginary probability in Eq.(\ref{eq:Imxp}) with the real probability in Eq.(\ref{eq:Rexp}). The classical limit of imaginary joint probabilities is then given by 
\begin{equation}
\mbox{Im}\left(\rho(x,p)\right) =  \frac{\hbar}{2} \frac{\partial^2}{\partial x \partial p} \mbox{Re}\left(\rho(x,p)\right).
\end{equation}
The appearance of $\hbar$ in this classical definition of imaginary probability indicates that, in the classical limit, the imaginary part will be much smaller than the real part. Oppositely, a joint probability can only be considered classical if the action given by the ratio of the joint probabilities and its second order derivative in $x$ and $p$ is sufficiently smaller than $\hbar$. 

Although the result above has been derived for the Kirkwood distribution in phase space, its generalization to the classical limit of high-dimensional discrete Hilbert spaces is straightforward. For slowly varying phases and amplitudes of $\langle a \mid b \rangle$, the corresponding expression can be obtained by replacing $\hbar=1/(2\pi|\langle p \mid x \rangle|^2)$ with $1/(2\pi |\langle a \mid b \rangle|^2)$. The result reads
\begin{equation}
\mbox{Im}\left(\rho(a,b)\right) =  \frac{1}{4 \pi |\langle a \mid b \rangle|^2} \frac{\partial^2}{\partial a \partial b} \mbox{Re}\left(\rho(a,b)\right).
\end{equation}
In general, the classical limit of imaginary probabilities can be represented by the gradients of the phase space distribution associated with general transformations of the parameters. The complex probabilities of quantum mechanics therefore represent a unification of statistics with the dynamics of transformations \cite{Hof11b}. The quantum of action defines the point at which the classical separation between dynamics and (static) information breaks down. At that point, it is necessary to include the topology of transformations in the definition of joint statistics, a task that is achieved most naturally by expressing quantum mechanics in terms of complex probabilities. 

%----

\section{On the empirical foundations of quantum determinism}
\label{sec:discuss}

Quantum determinism might have far reaching consequences for our understanding of quantum physics. However, the possibility of addressing seemingly counterintuitive properties of quantum mechanics in a new light may also cause new misunderstandings. In fact, the difficulty of identifying the precise physics behind useful mathematical concepts seems to be the very reason why there is so much fundamental disagreement on the proper interpretation of quantum mechanics. It may therefore be justified to take an extra sharp look at the physics that support and justify the use of complex joint probabilities.

As mentioned in the introduction, it is fundamentally impossible to perform two quantum measurements jointly. Nevertheless, all measurements can be performed in parallel, on separate representatives of the same system. That is why quantum theory does define the relations between completely different measurements, and physicists should try to make these relations as clear as possible. Unfortunately, previous constructions of joint probabilities such as the Wigner function or the one used in Feynman's explanation of quantum computation all exploited the ambiguity of partial measurement results, filling the gaps by convenient but necessarily arbitrary assumptions \cite{Har08,Fey82,Han96,Hof09}. It is therefore important to emphasize that the present approach is firmly rooted in the experimentally observable properties of quantum statistics.

Firstly, weak measurements can confirm complex joint probabilities directly. The only assumption used in the weak measurement is that the probabilities of the actual measurement outcomes of the weak measurement are proportional to the probability of the precise measurement result. Since this assumption clearly holds when no final measurement is performed (or when the final measurement confirms the weak measurement), it seems to be difficult to avoid the conclusion that the complex value obtained in a post-selected weak measurement represents the correct conditional probability. Moreover, weak measurement statistics can be observed directly in the back-action of strong measurements \cite{Joh07} and in the correlations between optimally cloned quantum systems \cite{Hof11b}. 

An essential point in the experimental evaluation of joint probabilities is the requirement of consistency. Weak measurement statistics require no implicit assumptions about correlations between different observables, since causality ensures that the weak measurement is not affected by the post-selection process, and the weakness of the measurement ensures that the final outcome is not influenced by the intermediate measurement. In contrast, the construction of the Wigner function from parallel measurements of linear combinations of $\hat{x}$ and $\hat{p}$ implicitly assumes that the eigenvalues of $\hat{x}+\hat{p}$ should be equal to the eigenvalues of $\hat{x}$ plus the eigenvalues of $\hat{p}$ - an assumption that is clearly inconsistent with operator algebra. 

Secondly, the joint complex probabilities discussed here are a natural mathematical choice based on the properties of operator algebras in Hilbert space. That is the reason why they were actually discovered long before their usefulness for the explanation of weak measurements and other paradoxical quantum statics were known. The definition of complex probabilities as expectation values of the products of two measurement operators is a simple representation of the ``AND'' operation in classical logic, where the truth value is also given by a product of the individual truth values. It therefore provides a natural expression for the joint validity of two quantum statements, without interpretational bias in favor of a specific type of measurement or physical system. 

One problem might be that quasi-probabilities have often been motivated by the assumption of quasi-realities, that is, by an understandable desire to return to some form of classical realism that defines objects in terms of completely measurement independent concepts. However, the present approach does the opposite: it shows that such an ersatz reality cannot be constructed from the mathematical objects that represent joint probabilities, and it explains how the measurement independent reality of classical physics can emerge as an approximation to the measurement dependent reality of quantum physics. Specifically, the functional relation between two measurement outcomes and a third measurement outcome that characterize the measurement independent determinism of classical physics are only approximations. Quantum mechanics does not provide a replacement for such classical determinisms. Instead, determinism is expressed in terms of statistical relations that should not be confused with the relative frequencies of classical statistics: a non-zero value of $p(c|a,b)$ does not mean that sometimes, the system is accidentally described by a,b, and c, but rather indicates that the separate frequencies of a, b, and c must be related to each other in a specific way, so that complete knowledge of the statistics of a and b means that we can determine the statistics of c as well. The complex values of these conditional probabilities are a strong indication that realist interpretations are not helpful. In fact, it seems that the present formulation of quantum statistics shows that determinism (and hence causality) does not require realism and actually contradicts realist assumptions in the quantum limit. 

An empirical interpretation of quantum mechanics requires that realism be restricted to the outcomes of actual measurements. In the context of this empirical realism, each individual system is characterized by its preparation and a single measurement outcome, where the specific form of both fully defines a context dependent reality accessible from the ``outside''. The complex joint probabilities discussed here apply to ensembles and indicate the statistical relations between different systems from the same source, measured in different ways. Thus, complex joint probabilities support and confirm the dependence of individual realities on the specific measurement context.

\section{Conclusions}

Complex joint probabilities provide a representation of quantum states in terms of any pair of observables with mutually overlapping eigenstates. Such states can never be measured jointly, but their statistical connection can be observed in weak measurements. The fundamental nature of this relation between incompatible quantum measurements is revealed when transformations between different joint probability representations are considered, since these relations describe how the deterministic relation between two measurements and a third measurement is described in quantum theory. The classical notion of completeness associated with phase space points thus survives in quantum mechanics. However, the complex probabilities associated with joint statements about non-commuting observables require a modification of classical determinism, so that the simultaneous assignment of measurement outcomes corresponding to measurement independent phase space points is impossible. Instead, determinism is expressed in terms of complex phases relating to the properties of phase space transformations. For sufficiently smooth phase space topologies, quantum determinism can be expressed by Gaussian distributions with imaginary variance. Thus, the differences between classical determinism and quantum determinism become relevant when the measurement resolution approaches or exceeds the imaginary variance of quantum determinism. 

The discussion above shows that the classical notion of reality emerges naturally from quantum contextuality when the measurement resolution is sufficiently low. The idea of a measurement independent reality ``out there'' may therefore reflect a reasonable approximation, similar to the assumption of a flat space time in the absence of strong gravitational fields. Importantly, the lack of measurement independent realities can now be explained in terms of precise deterministic relations between the different possible measurements. Hilbert space thus provides a well-defined quantum limit of phase space topologies. In the context of time evolution and causality, this means that a single pair of observables determines the complete history of a quantum object. However, this history cannot be described by assigning a time-dependent value to a specific property, since such an assignment corresponds to simultaneous measurements of multiple non-commuting properties. Instead, quantum determinism only provides precise statements about the relation between the measurement statistics obtained for different representatives of the same source measured at different times. 

Even from a merely technical viewpoint, quantum determinism should proof useful by providing a consistent measurement-based description of quantum mechanics. The reformulation of Hilbert space concepts in terms of statistical expressions may be particularly useful in the analysis of the quantum information content of states as suggested by related approaches to quantum statistics that contributed to the motivation for the present work \cite{Bru99,Law02,Pat09,Dia05,Ren04,Fuc09,Med11,Rau09}. From my own perspective, however, the most surprising aspect of the present work is the possibility of defining deterministic relations between different measurements that are independent of the assignment of simultaneous values to the measurements and actually contradict such assignments in all precisely defined cases. This means that there is actually much less freedom in the interpretation of quantum mechanics than previously thought. In particular, quantum determinism appears to introduce a complete definition of the fabric of empirically accessible reality, representing an entirely new framework for all experimentally accessible aspects of quantum physics. Once the topology of quantum determinism is fully understood, it may finally be possible to explain quantum mechanics entirely in terms of empirical concepts, without the need for postulates in the form of unmotivated mathematical abstractions. 

\section*{Acknowledgment}
Part of this work has been supported by the Grant-in-Aid program of the Japanese Society for the Promotion of Science, JSPS.

\end{document}